\documentclass[11pt]{article}

\usepackage{graphicx}
\usepackage{dcolumn}
\usepackage{amsmath}

\input pubboard/babarsym
\newcommand{\deltae}{\ensuremath{\Delta E}}

\def\Dp      {\ensuremath{D^+}}
\def\Dm      {\ensuremath{D^-}}
\def\Dstar   {\ensuremath{D^{*}}}
\def\Dstarp  {\ensuremath{D^{*+}}}
\def\Dstarm  {\ensuremath{D^{*-}}}

\def\rhop    {\ensuremath{\rho^+}}
\def\rhom    {\ensuremath{\rho^-}}
\def\aonep   {\ensuremath{a_1^+}}

\providecommand{\hips}{\mbox{\ensuremath{\hbar\ps^{-1}}}}
\newcommand{\BABARPubYear}    {01}

\newcommand{\BABARConfNumber} {23}

\newcommand{\SLACPubNumber} {8913}

\def\figurebox#1#2#3{%
    \def\arg{#3}%
    \ifx\arg\empty
    {\hfill\vbox{\hsize#2\hrule\hbox to #2{\vrule\hfill\vbox to #1{\hsize#2\vfill}\vrule}\hrule}\hfill}%
    \else
    {\hfill\epsfbox{#3}\hfill}%
    \fi}

\long\def\inst#1{\par\nobreak\kern 4pt\nobreak
    {\it #1}\par\vskip 10pt plus 3pt minus 3pt}

\begin{document}
{\pagestyle{empty}

\begin{flushright}
\babar-CONF-\BABARPubYear/\BABARConfNumber \\
SLAC-PUB-\SLACPubNumber \\
July, 2001 \\
\end{flushright}

\par\vskip 5cm

\begin{center}
Measurement of the {\boldmath $\Bz\Bzb$} Oscillation Frequency in
Hadronic {\boldmath \Bz} Decays 
\end{center}
\bigskip

\begin{center}
\large The \babar\ Collaboration\\
\mbox{ }\\
July 11, 2001
\end{center}
\bigskip \bigskip

\begin{center}
\large \bf Abstract
\end{center}
$\BzBzb$ flavor oscillations have been studied in 20.7 \invfb\ of
\epem\ annihilation data collected in 1999 and 2000 with the
\babar\ detector at center-of-mass energies near the \FourS\ resonance.
The event sample consists of one \Bz\ meson fully reconstructed in a hadronic decay mode, 
while the flavor of the recoiling \Bz\ in the event is determined with a tagging 
algorithm that exploits the correlation between the flavor of the 
heavy quark and the charges of its decay products. 
By fitting the time development of the
observed mixed and unmixed final states, 
the $\BzBzb$ oscillation frequency,
\deltamd, is determined to be $0.519\pm 0.020\pm 0.016\,\hbar\ps^{-1}$. 

\vfill
\begin{center}
Submitted to the International Europhysics 
Conference on High Energy Physics, \\
7/12---7/18/2001, Budapest, Hungary
\end{center}

\vspace{1.0cm}
\begin{center}
{\em Stanford Linear Accelerator Center, Stanford University, 
Stanford, CA 94309} \\ \vspace{0.1cm}\hrule\vspace{0.1cm}
Work supported in part by Department of Energy contract DE-AC03-76SF00515.
\end{center}
}
\newpage

\begin{center}
\small

The \babar\ Collaboration,
\bigskip

B.~Aubert,
D.~Boutigny,
J.-M.~Gaillard,
A.~Hicheur,
Y.~Karyotakis,
J.~P.~Lees,
P.~Robbe,
V.~Tisserand
\inst{Laboratoire de Physique des Particules, F-74941 Annecy-le-Vieux, France }
A.~Palano
\inst{Universit\`a di Bari, Dipartimento di Fisica and INFN, I-70126 Bari, Italy }
G.~P.~Chen,
J.~C.~Chen,
N.~D.~Qi,
G.~Rong,
P.~Wang,
Y.~S.~Zhu
\inst{Institute of High Energy Physics, Beijing 100039, China }
% ...

\end{center}\newpage

\section{Introduction}
\label{sec:Introduction}

In the Standard Model,
$\BzBzb$~\cite{conjugate} mixing occurs through second-order weak diagrams
involving the exchange of up-type quarks, 
with the top quark contributing
the dominant amplitude.  A measurement of \deltamd, the difference between 
the mass eigenstates of the \Bz--\Bzb system, is therefore
sensitive to the magnitude of the Cabibbo-Kobayashi-Maskawa matrix~\cite{CKM} element
$V_{td}$.
At present the sensitivity to $V_{td}$ is not limited by
experimental precision on \deltamd, but by other
uncertainties in the calculation, in particular the quantity
$f_B^2 B_B$, where $f_B$ is the \Bz\ decay constant, and 
$B_B$ is the so-called bag factor, 
representing  the strong interaction matrix elements.

The phenomenon of particle--anti-particle mixing 
in the neutral \B\ meson system was first observed almost fifteen years
ago~\cite{B0mix}.  
The oscillation frequency
has been extensively studied with both time-integrated  and 
time-dependent techniques~\cite{PDG2000}.  

In this paper we present a measurement of time-dependent mixing 
based on a sample of $20.7\invfb$ of data recorded
at the \FourS\ resonance with the \babar\ detector at the Stanford Linear
Accelerator Center.
At the PEP-II asymmetric-energy $e^+e^-$ collider, resonant production of
the \FourS\  provides a 
copious source of $\BzBzb$ pairs moving along the beam axis ($z$ direction)   
with a Lorentz boost of $\beta\gamma = 0.56$.
The typical separation between the two neutral $B$  decay vertices is
$\left<|\deltaz|\right> \approx \beta\gamma c \tau_{\Bz} = 260$\mum, where $\tau_{\Bz} =
1.548\pm 0.032$\ps\ is the \Bz\ lifetime~\cite{PDG2000}.

\section{Analysis method}
\label{sec:Analysis}
The \BzBzb\ mixing probability 
is a function of \deltamd\ and the 
time difference between the $B$ decays, $\deltat \simeq \deltaz/\beta\gamma c$:
\begin{eqnarray}
Prob(\Bz\Bzb \rightarrow \Bz\Bz, \Bzb\Bzb) & \propto & {\rm{e}}^{-|\deltat|/\tau_{\Bz}}(1 - \cos \deltamd \deltat) \nonumber \\
Prob(\Bz\Bzb \rightarrow \Bz\Bzb )         & \propto & {\rm{e}}^{-|\deltat|/\tau_{\Bz}}(1 + \cos \deltamd \deltat) \nonumber
\label{eq:mix} 
\end{eqnarray}
resulting in a time-dependent probability to observe {\em mixed}, $\Bz\Bz$ and $\Bzb\Bzb$, or {\em unmixed},
$\BzBzb$, pairs produced in
\FourS\ decay.  The effect can be measured by reconstructing
one \B\ in a flavor eigenstate, referred to as $B_{rec}$, while using the remaining particles from
the decay of the other \B, referred to as $B_{tag}$, to identify, or ``tag'', its flavor.
The charges of identified leptons and kaons are the primary indicators, 
but other information
in the event can also be used to identify the flavor of $B_{tag}$, resulting
in a total of four non-overlapping tagging categories.
The tagging algorithm used in this analysis is identical to that
employed for \CP\ violation studies, 
in which one \B\ is fully reconstructed in a
\CP\ eigenstate~\cite{BabarPub0101}.  

If the flavor tagging were perfect, 
the asymmetry as a function of $\Delta t$
\begin{equation*}
A_{mixing}(\deltat) = \frac{N_{unmixed}-N_{mixed}}{N_{unmixed}+N_{mixed}} 
\label{eq:asym}
\end{equation*}
would be a cosine with unit amplitude.
However, the tagging algorithm incorrectly identifies the flavor of $B_{tag}$ with a 
probability $\mistag_i$ for the $i^{th}$ tagging category.  This mistag rate reduces the
amplitude of the oscillation by a
factor $(1 -2\mistag_i)$. 
A simultaneous fit to the mixing frequency
and its amplitude allows the determination of both \deltamd\
and the mistag rates, $\mistag_i$.  

Neglecting any background contributions, 
the probability density functions (PDFs) 
for the mixed $(+)$ and unmixed $(-)$ events, ${\cal H}_{\pm, sig}$, can be expressed as
the convolution of the decay distribution for the $i^{th}$ tagging category
\begin{equation*}
h_\pm(\deltat; \deltamd, \mistag_i)  = \frac{{\rm e}^{ -\left| \deltat \right|/\tau_{\Bz}}}{4\tau_{\Bz}}
\left[ 1 \mp (1-2\mistag_i)\cos{ \deltamd \deltat } \right],
\end{equation*}
with a time resolution
function ${\cal R}(\deltat-\deltat_{true};\hat {a}_i)$,
\begin{equation*}
\label{eq:pdf}
{\cal H}_{\pm, sig}(\deltat; \deltamd, \mistag_i, \hat {a}_i ) =
h_\pm(\deltat_{true}; \deltamd, \mistag_i ) 
\otimes 
{\cal R}( \deltat-\deltat_{true} ; \hat {a}_i ).
\end{equation*}
where $\hat {a}_i$ are the parameters of the resolution function.
A log-likelihood function is then constructed from the sum of $\log{\cal {H}}_{\pm, sig}$ over all mixed
and unmixed events, and 
over the different tagging categories.

The log-likelihood is maximized to extract the mixing parameter \deltamd and,
simultaneously, the mistag rates, $\mistag_i$.
The correlation among these parameters is small, because
the rate of mixed events at low values of \deltat, where
the \BzBzb\ mixing probability is small, is principally 
governed by the mistag rate. Conversely, the sensitivity 
to \deltamd\ increases at larger values of \deltat;
for \deltat\ near twice the lifetime, half of
the \Bz\ mesons will have oscillated.

\section{The \babar\ detector}
\label{sec:babar}

The \babar\ detector is a $4\pi$ charged and neutral spectrometer
described in more detail elsewhere~\cite{BabarPub0108}.
Charged particles are detected and their momenta  measured by a combination of 
a 40-layer drift chamber (DCH) and a five-layer silicon vertex tracker (SVT) embedded
in a 1.5-T solenoidal magnetic field. 
Decay vertices 
are typically reconstructed with a resolution along the boost direction of 
65\mum\ for fully reconstructed $B$ mesons.
A ring imaging Cherenkov
detector, the DIRC, is used for charged hadron identification.
A finely segmented CsI(Tl)
electromagnetic calorimeter (EMC) is used to 
detect photons and neutral hadrons, and also for electron
identification. 
The iron flux return (IFR) is segmented and instrumented with
multiple layers of resistive plate chambers
for the identification of muons and long-lived neutral hadrons. 

\section{Event selection and $B$ reconstruction}
\label{sec:evntselection}
The analysis uses a sample of multihadron events, which are
selected by demanding a minimum of three reconstructed charged tracks
and a total charged and neutral energy greater than 4.5\gev\ in the fiducial region of the detector,
and a reconstructed event vertex within 0.5\cm\ of the measured
interaction point~\cite{BabarPub0108} in the plane transverse to the beamline.

Electron candidates must satisfy a cut on the ratio of calorimeter
energy to track momentum of $0.88 < E/p < 1.3$, a cluster shape
consistent with an electromagnetic shower, and DCH \dedx\ and DIRC
Cherenkov angle consistent with an electron.

Muon candidates must satisfy requirements on the number of
interaction lengths of IFR iron penetrated of $N_{\lambda} > 2.2$, on
the difference in the measured and expected interaction lengths
penetrated of $N_{\lambda}^{exp} - N_{\lambda} < 1$, on the position match
between the extrapolated DCH track and the IFR hits, and on the average
and spread of the number of IFR hits per layer.

Pairs of photons in the EMC with energy above 30\mev
are constrained to the known \piz\ mass if they are 
within $\pm 20$\mevcc\ of the nominal
invariant mass~\cite{PDG2000}, and their summed energy is greater than 200\mev. 

$\KS\to\pip\pim$ candidates are required to have an
invariant mass between 462
and 534\mevcc, and a $\chi^2$ probability for the vertex fit of
greater than 0.1\%. 
The transverse flight distance of the \KS\ candidate
from the primary event vertex must be greater than 2\mm.

\Dzb\ candidates are identified in the 
decays channels $\Kp\pim$, $\Kp\pim\piz$, $\Kp\pip\pim\pim$ and $\KS
\pip\pim$. \Dm\ candidates are selected in the $\Kp \pim \pim$
and $\KS\pim$ modes.
Kaons from \Dm\ decays and charged daughters from \Dzb\to\Kp\pim are required
to have a momentum greater than 200\mevc. All other charged \Db\ daughters
are required to have a momentum greater than 150\mevc.
For $\Dzb\to\Kp \pim \piz$, we only reconstruct the dominant resonant
mode $\Dzb\to\Kp \rhom$, followed by $\rhom\to\pim \piz$. The $\pim \piz$ mass is
required to lie within $\pm 150$\mevcc\ of the nominal $\rho$ mass~\cite{PDG2000} and
the angle between the \pim\ and \Dzb\ in the $\rho$ rest frame, 
$\theta^*_{\Dz\pi}$, must satisfy $|\cos \theta^*_{\Dz \pi}| > 0.4$.
\Dzb\ and \Dm\ candidates are required
to have momentum greater than 1.3\gevc\
in the \FourS\ frame, an invariant mass within 
$\pm 3\sigma$ of the nominal value~\cite{PDG2000}
and a $\chi^2$ probability of the topological vertex fit 
greater than 0.1\%.  A mass constraint is applied to selected
\Dzb candidates.

\Dstarm\ candidates are formed by combining a \Dzb\ and a pion with
momentum less than 450\mevc.
The soft pion is constrained to originate from the beamspot
when the \Dstarm\ vertex is computed. 
Those \Dstarm\ candidates with $m(\Dzb\pim)-m(\Dzb)$ lying
within $\pm 2.5\sigma$ of the nominal value~\cite{PDG2000} are selected, 
where $\sigma=1.1$\mevcc\ for $\Dzb\to\Kp \pim \piz$ mode 
and 0.8\mevcc\ for all other modes.

$\jpsi \to \epem$ or $\mu^+ \mu^-$ candidates must have at least one 
decay product
positively identified as an electron or muon.
Electron candidates outside the calorimeter acceptance 
must have DCH \dedx\ information consistent with that for an electron.
The second muon candidate, if within the acceptance of the calorimeter, must be
consistent with being a minimum ionizing particle.
$\jpsi$ candidates are required to lie in the invariant mass interval 
2.95 (3.06) to 3.14\gevcc\ for the $\epem$ ($\mu^+\mu^-$) channel.

\Bz\ candidates in the flavor eigenstate decay modes $D^{(*)-}\pip/\rho^+/a_1^+$  
are formed by combining a \Dstarm\ or \Dm\ candidate
with a \pip, \rhop\ $(\rhop\to\pip\piz)$ or \aonep\ $(\aonep\to\pip\pim\pip)$;
likewise
$\Bz\to\jpsi\Kstarz$ candidates are reconstructed
from combinations of \jpsi\ candidates with a \Kstarz\ $(\Kstarz\to\Kp\pim)$.  

For $\Bz\to\Dstarm
\rhop$, the \piz\ from the \rhop\ decay is required to have an
energy greater than 300\mev. For $\Bz\to\Dstarm\aonep$,
the \aonep\ is reconstructed by combining three charged
pions, with invariant mass in the range 1.0 to 1.6\gevcc\ and a
$\chi^2$ probability of the vertex fit of the \aonep\ candidate of
greater than 0.1\%.
For most \Bz\ modes, no particle identification or only a loose requirement
is enough to achieve reasonable signal purities. 

Continuum background is rejected by requiring the normalized second Fox-Wolfram moment~\cite{fox} be less than 0.5.
Further suppression is achieved by a mode-dependent
restriction on the 
angle, $\theta_{th}$, between the thrust axes of decay products from $B_{rec}$ and $B_{tag}$ respectively
in the \FourS\ frame.

\Bz\ candidates are identified with
the difference \deltae\ between the 
energy of the candidate 
and the beam energy $E^{\rm cm}_{\rm beam}$ in the center-of-mass frame and
the beam-energy substituted mass 
$\mes=\sqrt{{(E^{\rm cm}_{\rm beam})^2}-(p_B^{\rm cm})^2}$.
Those candidates with $\mes>5.2$\mevcc\
and $\Delta E=0$ within $\pm 2.5$ standard deviations
(typically $\left| \Delta E \right| < 40$\mev) are selected. If there is more than one candidate
satisfying these conditions 
only the one with the smallest \deltae\ is retained.
Finally, a topological vertex fit of the candidate must converge.

\begin{figure}[!tbh]
\begin{center}
 \includegraphics[width=0.90\linewidth]{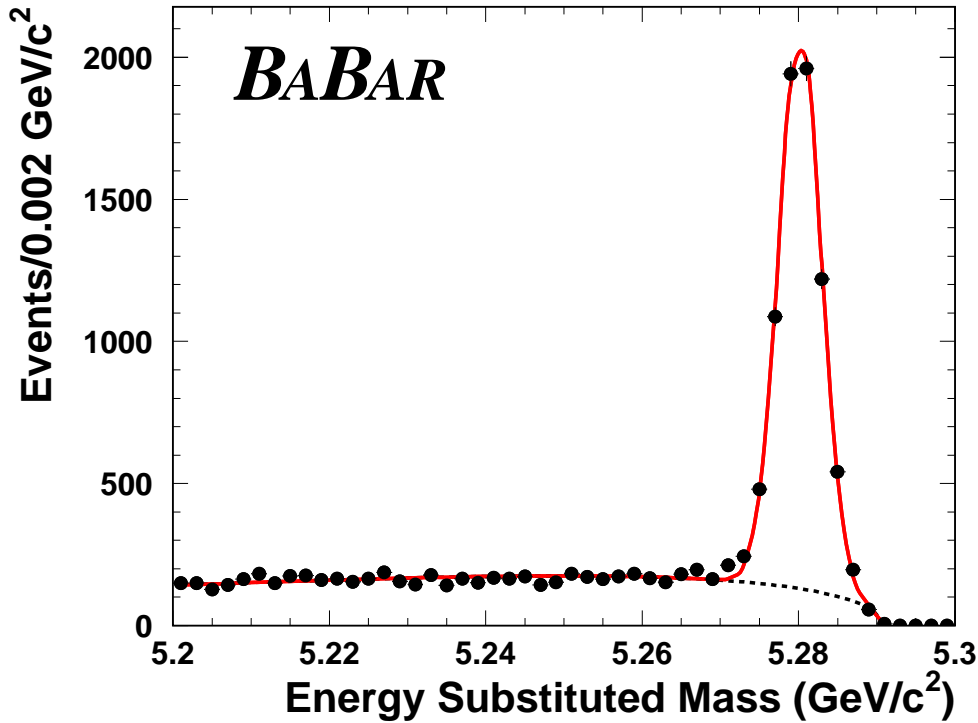}
\end{center}
\caption{Distribution of \mes\ for all \Bz\ hadronic 
candidates decaying into flavor eigenstate modes.
\label{fig:hadronicb0}}
\end{figure}

\section{Decay time difference determination}
\label{sec:decaytime}
The decay time difference, \deltat,
between \B\ decays is determined from the
measured separation along the $z$ axis between the reconstructed $B=B_{rec}$ and 
flavor-tagging decay $B=B_{tag}$ vertices $\Delta z = z_{rec}-z_{tag}$.
This $\Delta z$ is then converted into \deltat using the known \FourS\ boost
and correcting on an event-by-event basis for the direction of the \B\ mesons 
with respect to the $z$ direction in the \FourS\ frame.
The resolution of the \deltat\ measurement
is dominated by the $z$ resolution of the tagging vertex.
The $B_{tag}$ decay vertex uses all
tracks in the event except those incorporated in $B_{rec}$. 
An additional constraint is provided by including a calculated
$B_{tag}$ production point and three-momentum, determined
from the three momentum of the $B_{rec}$ candidate, its decay vertex,
and the average position of the interaction point and the \FourS\ boost. 
Reconstructed $\KS$ or $\Lambda$
candidates are used as input to the fit in place
of their daughters in order to reduce
bias due to long-lived particles.
Tracks with a large contribution to the $\chi^2$
are iteratively removed from the fit, 
until all remaining tracks have a reasonable fit probability 
or all tracks are removed. Only those candidates
with $|\Delta z|<3.0$\mm\ and $\sigma_{\Delta z}<400$\mum\ are retained.
The distribution of \mes\ for the surviving candidates is shown
in Fig.~\ref{fig:hadronicb0} along with a fit with a Gaussian distribution
for the signal and the ARGUS function~\cite{ARGUS_bkgd} for
the background.
At this point the sample contains $6662 \pm 93$ signal events,
with a purity, defined in the region $\mes>5.27\gevcc$,
that varies between 80--95\% depending on the \Bz\ decay mode.

In the likelihood, the time resolution function 
can be approximated by a sum of three Gaussian distributions with different means and widths,
\begin{equation*}
{\cal R}( \deltat ; \hat {a}_i ) =  \sum_{k=1}^{3} 
{ \frac{f_k}{\sigma_k\sqrt{2\pi}} {\rm e}^{ 
-(\deltat-\deltat_{true}-\delta_{k,i} )^2/2{\sigma_k}^2} },
\end{equation*}
where, for the core and tail Gaussians, the widths
$\sigma_{1,2}=S_{1,2}\times\sigma_{\deltat}$ are the scaled event-by-event 
measurement error, $\sigma_{\deltat}$, derived from the vertex fits. 
The third Gaussian, with a fixed width of $\sigma_3=8$\ps, accounts for 
less than 1\% of outlier events with
incorrectly reconstructed vertices. 
A separate core bias, $\delta_{1,i}$, is allowed 
for each tagging category $i$ to account for small
shifts due to inclusion of charm decay products in the tag vertex, 
while a common bias $\delta_2$ is used for the
tail component. 
The tail and outlier fractions and the scale factors 
are assumed to be the same for all decay modes,
since the precision of the $B_{tag}$ vertex dominates \deltat. This
assumption is confirmed by Monte Carlo simulation studies.

\section{Flavor tagging}
\label{sec:flavtag}
After the daughter tracks of the reconstructed $B$ are removed,
the remaining tracks are analyzed to determine the flavor of the
$B_{tag}$, and this ensemble is assigned a tag flavor, either \Bz\ or \Bzb.  
For this purpose, we use the flavor tagging information carried by primary leptons from
semileptonic $B$ decays, charged kaons, soft pions from \Dstar\ decays, and
more generally by high momentum charged particles to uniquely assign each event
to a tagging category. 
The effective tagging efficiency
$Q_i = \eps_i (1-2\mistag_i)^2$, where $\eps_i$ is the fraction of events assigned to
category $i$, is used
as the basis for optimization of
category selection criteria. The statistical error on
\deltamd\ is proportional to $1/\sqrt{Q}$, where $Q = \sum {Q_i}$.
%

Events are assigned to the {\tt Lepton} category
if they contain an identified lepton
with a center-of-mass momentum greater than 
1.0 or 1.1\gevc\ for electrons and muons, respectively. 
The momentum requirement selects mostly primary leptons by suppressing
opposite-sign leptons from semileptonic charm decays. 

Kaons are identified with a neural network based on the  
likelihood ratios in the SVT
and DCH, derived from \dedx\ measurements, and in the DIRC,
calculated by comparing individual photomultiplier hits with the
expected pattern of Cherenkov light for either kaons or pions.
The charges of all identified kaons are summed, and
if $\sum Q_K \ne 0$ the event is assigned to the {\tt Kaon} category. 

The final two categories involve a multivariable analysis using a 
neural network, which is trained
to identify primary leptons, kaons, and soft pions, and
the momentum and charge of the track with the maximum center-of-mass
momentum.  Depending on the value of the output variable from the neural net, events are given
a \Bz\ or \Bzb\ tag and assigned to the mutually exclusive 
categories {\tt NT1} (more certain tags) or {\tt NT2}
(less certain tags). About 30\% of all
events are assigned to no tagging category and are excluded from the analysis.

Tagging assignments for events are made mutually exclusive by the hierachical
use of the tagging categories. Events with a {\tt Lepton} tag and no conflicting {\tt Kaon}
tag use the {\tt Lepton} category. If no {\tt Lepton} tag exists, then the {\tt Kaon} category
is used, if a tag exists. Otherwise the two neural network categories are used.
The number of tagged events per category is given in Table~\ref{tab:HadronicBYield}.
In total, there are $4538\pm 75$ tagged signal events.

\begin{table}[!htb]
\caption{
Event yields for the different tagging categories used in this analysis, as 
obtained from fits to the \mes\ distributions after all selection 
requirements.  The purity is quoted for 
$\mes >5.270 \mevcc$.} 
\begin{center}
\begin{tabular}{lccc} \hline\hline
Category     & Tagged       & Efficiency (\%) & Purity (\%) \\ \hline
{\tt Lepton} & $ 754\pm 28$ & $11.3\pm 0.4$    & $97.1\pm 0.6$  \\
{\tt Kaon}   & $2317\pm 54$ & $34.8\pm 0.6$    & $85.2\pm 0.8$  \\
{\tt NT1}    & $ 556\pm 26$ & $ 8.3\pm 0.3$    & $88.7\pm 1.5$  \\ 
{\tt NT2}    & $ 910\pm 36$ & $13.7\pm 0.4$    & $83.0\pm 1.3$  \\ \hline
Total        & $4538\pm 75$ & $68.1\pm 0.9$    & $86.7\pm 0.5$  \\ \hline\hline
\end{tabular}
\end{center}
\label{tab:HadronicBYield}
\end{table}

\section{Background PDF}

In the presence of backgrounds, the probability distribution functions 
${\cal H}_{\pm, sig}$
must be extended to include a 
term for each significant background source, which are
allowed to differ for each tagging category:
\begin{equation}
{{\cal H}_{\pm,i}} =
 f_{i,sig}{\cal H}_{\pm, sig} + 
 \sum_{k={bkgd}} f_{i,k} {\cal{B}}_{\pm,i,k}(\deltat;\hat b_{\pm,i,k}) \nonumber
\end{equation}
where
the background PDFs, ${\cal{B}}_{\pm,i,k}$, provide an empirical 
description for the \deltat\ distribution of the background events in the sample.  
The fraction of background events for each source and tagging category is given by  
$f_{i,k}$, while $\hat b_{\pm,i,k}$ are parameters used to characterize each
source of background by tagging category for mixed and unmixed events. 

The probability that a \Bz\ candidate is a signal or a background
event is determined from a separate fit to the observed \mes\ distributions of $B_{rec}$ candidates in each
of the four tagging categories.
We describe the \mes\ shape with a single Gaussian distribution ${\cal S}(\mes)$
for the signal and an ARGUS parameterization ${\cal A}(\mes)$ for
the background. Based on this fit, the event-by-event signal
probabilities $f_{sig, i}$ are given by  
\begin{equation*}
f_{i, {\rm sig}}(\mes) =  \frac{{\cal S}(\mes)}{{\cal S}(\mes)+{\cal A}(\mes)}
\label{eq:SigProb}
\end{equation*}
The sum of signal and background fractions is forced to unity. 

The \deltat\ distributions of the combinatorial
background are assumed to be described with a zero lifetime component and a 
non-oscillatory component with non-zero lifetime. 
We fit for separate resolution function
parameters for the signal and the background in order
to minimize correlations
of the time structure between background and signal. Candidates with
low signal probability, {\em i.e.,} in the \mes\ sideband region below 5.27\gevcc,
dominate the determination of these background parameters.

\begin{table}[!htb]
\begin{center}
\caption{Results from the likelihood fit to the \deltat\ 
distributions of the tagged hadronic \Bz\ decays. \deltamd\ and 
the mistag rates include small corrections corresponding to the 
difference between the generated and reconstructed values in simulated signal events.} 
\label{tab:result-likeli}
\begin{tabular}{lclc}
\hline\hline
Parameter                & Value         &
Parameter                & Value         \\ \hline
\deltamd (\hips)\        & $0.519 \pm 0.020$ & $\delta_{1,{\tt Lepton}}(\ps)$         & $+0.11\pm 0.07$   \\
$\mistag_{\tt Lepton}$   & $0.085\pm 0.018$  & $\delta_{1,{\tt Kaon}}(\ps)$           & $-0.20\pm 0.05$   \\
$\mistag_{\tt Kaon}  $   & $0.167\pm 0.014$  & $\delta_{1,{\tt NT1}}(\ps)$            & $+0.01\pm 0.09$   \\
$\mistag_{\tt NT1}   $   & $0.195\pm 0.026$  & $\delta_{1,{\tt NT2}}(\ps)$            & $-0.20\pm 0.08$   \\
$\mistag_{\tt NT2}   $   & $0.326\pm 0.024$  & $\delta_2(\ps)$                       & $-1.2^{+0.9}_{-1.6}$    \\
$S_1$                    & $1.42^{+0.08}_{-0.09}$& $f_2$                            & $0.032^{+0.03}_{-0.02}$ \\
$S_2$                    & $5.5^{+2.0}_{-1.6}$   & $f_3$                            & $0.001\pm 0.004$        \\ \hline\hline
\end{tabular}
\end{center}
\end{table}

\section{Extraction of \deltamd}
\label{sec:deltamd}

The value of \deltamd\ is extracted from the tagged flavor-eigenstate \Bz\ sample with an unbinned
maximum likelihood fit
involving a total of 34 parameters, including \deltamd.
The value of \deltamd\ was kept hidden throughout the analysis until
the systematic errors were finalized,
in order to eliminate possible experimenter's bias.

\begin{figure}[!htb]
\begin{center}
  \includegraphics[width=0.90\linewidth]{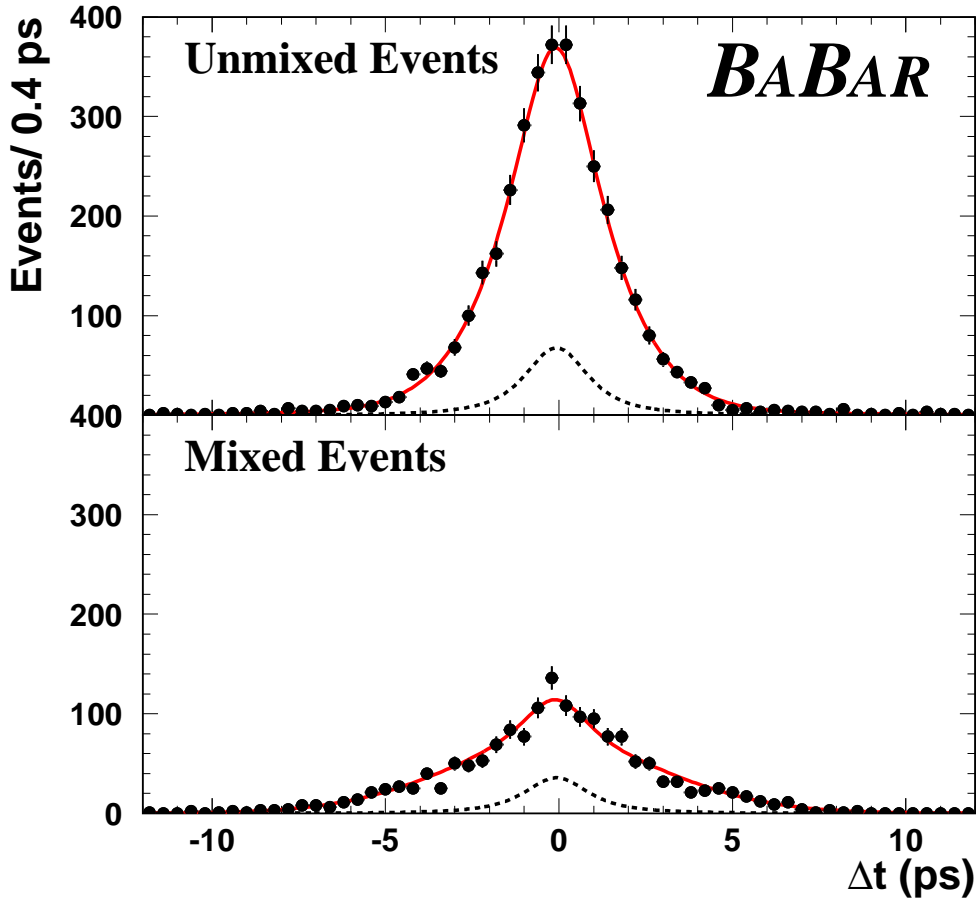}
  \caption{\deltat\ distributions in data for the selected mixed and unmixed
    tagged hadronic \Bz\ decays ($\mes>5.27$\gevcc), with overlaid the projection
    of the likelihood fit. The background contribution is indicated by the
    dashed curve.
    }
\label{fig:deltat}
\end{center}
\end{figure}

\begin{figure}[!htb]
\begin{center}
  \includegraphics[width=0.90\linewidth]{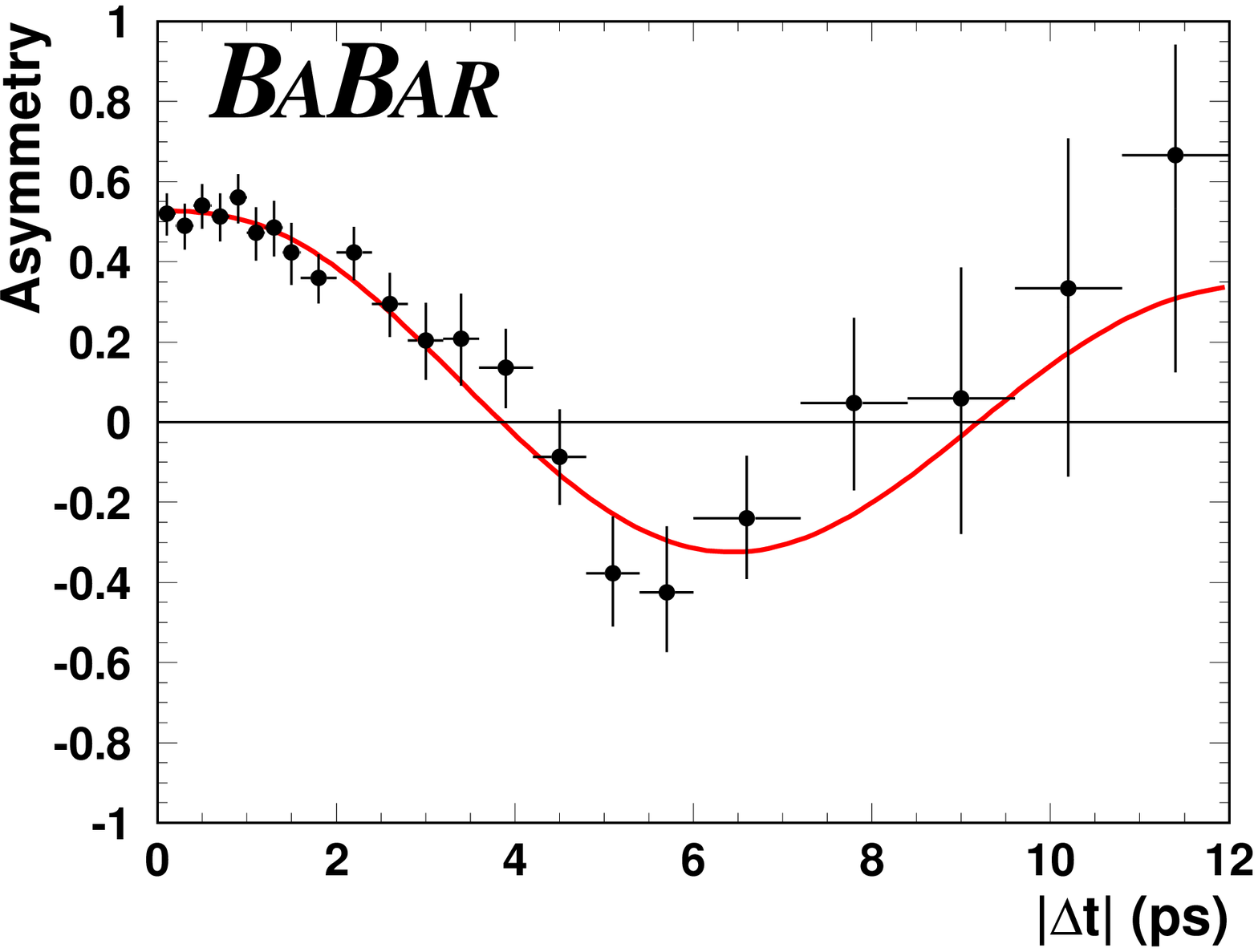}
  \caption{The 
    time-dependent asymmetry ${\cal A}_{mix}(\deltat)$ between unmixed and
    mixed events as a function of $|\deltat|$.
    }
\label{fig:asymm}
\end{center}
\end{figure}

The results from the likelihood fit to the tagged \Bz\ sample are summarized in
Table~\ref{tab:result-likeli}. 
The probability of obtaining a likelihood smaller than the one observed 
is determined to be $28\%$ from a large number of simulated experiments,
each generated according to the parameters obtained from the fit.
The measured value for \deltamd\ is:
\begin{equation*}
\deltamd = 0.519\pm 0.020 ({stat}) \pm 0.016 ({syst})\,\hbar\ps^{-1}.
\end{equation*}
where the sources of systematic error
are discussed below.
The observed distribution of mixed and unmixed events 
and the asymmetry, ${\cal A}_{mix}$
are shown in Fig.~\ref{fig:deltat} and \ref{fig:asymm}
as a function of \deltat\ along with
projections of the likelihood fit result.

\section{Systematic studies and cross checks}
\label{sec:Systematics}

The conversion of \deltaz\ to \deltat\ introduces
a systematic uncertainty ($\pm 0.007\,\hbar\ps^{-1}$)
due to the limited knowledge of the PEP-II boost, 
the $z$ length scale of \babar\ (determined from 
secondary interactions in a beam pipe section of known
length) and the $B_{\rm rec}$ momentum vector in the
\FourS\ frame.

The signal \deltat\ resolution parameters are determined directly from data
by the fit, contributing $\pm 0.006$ in quadrature to the statistical error. 
Residual uncertainties ($\pm 0.005\,\hbar\ps^{-1}$),
are attributed to the choice of the parameterization,
its description of the outliers, and the capability of the
resolution model to deal with various plausible misalignment scenarios 
applied to the Monte Carlo simulation.

The parameters of the background \deltat\ distribution
are left free in the likelihood fit, but systematic errors
($\pm 0.005\,\hbar\ps^{-1}$),
are introduced by the residual uncertainty from the
\mes fit used to determine the signal probability,
the assumed parameterization of the background \deltat\
distributions and resolution function, and the small amount 
of correlated \Bu\ background remaining in the sample.

Finally, statistical limitations of Monte Carlo validation tests ($\pm 0.004\,\hbar\ps^{-1}$),
the full size of a (negative) correction obtained from Monte Carlo ($\pm 0.009\,\hbar\ps^{-1}$),
and the variation of the fixed \Bz\ lifetime (due to its negative correlation with \deltamd)
within known errors~\cite{PDG2000} ($\pm 0.006\,\hbar\ps^{-1}$) contribute.
A summary
of these sources of systematic error for the hadronic \Bz\ sample
is shown in Tables~\ref{tab:systematics}.

Various checks on the consistency of the result were performed by
splitting the data in sub-samples according to several key variables, including
(but not limited to)
the decay modes of $\B_{rec}$, the tagging category of $\B_{tag}$,
and the flavor of either $\B_{rec}$ or $\B_{tag}$. The value of \deltamd\
was found to be consistent for all sub-samples.

\begin{table}[htb]
\begin{center} 
 \caption{Systematic uncertainties for \deltamd.\label{tab:systematics}}
  \begin{tabular}{lc}\hline\hline
    Source  & $\delta\Delta m_d$ [$\hbar$ ps$^{-1}$] \\
    \hline
    Beamspot               &         \\
    \hline 
    \ \ position and size  &  $0.001$ \\
    \deltaz\ to \deltat\ conversion &  \\
    \hline
    \ \ PEP-II boost       & $0.003$ \\
    \ \ $z$ scale          & $< 0.005$ \\
    \ \ method             & $0.004$ \\
    \deltat\ resolution    &         \\
    \hline
    \ \ outliers           & $0.002$ \\
    \ \ parameterization   & $0.003$ \\
    \ \ SVT alignment      & $0.004$ \\
    Backgrounds            &       \\
    \hline 
    \ \ \deltat\ model     & $0.001$ \\
    \ \ Resolution parameterization & $0.003$ \\
    \ \ Fractions          & $0.003$ \\
    \ \ Correlated         & $0.002$ \\
    Monte Carlo            &          \\
    \hline
    \ \ statistics         & $0.004$ \\
    \ \ correction         & $0.009$\\
    \hline
    \Bz\ lifetime          & $0.006$ \\
    \hline
    Total Systematic Error       & $0.016$ \\
        \hline\hline
   \end{tabular}
\end{center}
\end{table}

\section{Summary}
\label{sec:Summary}

We have measured the value for \deltamd\ to be:
\begin{equation*}
\deltamd = 0.519\pm 0.020 ({stat}) \pm 0.016 ({syst})\,\hbar\ps^{-1}.
\end{equation*}
This result is one of the best single measurements available and
is consistent with the current world average~\cite{PDG2000}.
Moreover, the error on \deltamd\ is still dominated by the statistical size
of the reconstructed \Bz\ sample, leaving substantial room for further
improvement as more data is accumulated at \babar. The measurement
shares the same flavor-eigenstate sample as used for the determination
of $\sin\! 2\beta$. Thus, it
provides an essential
validation for the reported $\sin\! 2\beta$
result~\cite{BabarPub0101}
and, in particular, the mistag rates that appear as coefficients
of the mixing asymmetry. 

\section{Acknowledgments}
\label{sec:Acknowledgments}

We are grateful for the 
extraordinary contributions of our \pep2\ colleagues in
achieving the excellent luminosity and machine conditions
that have made this work possible.
The collaborating institutions wish to thank 
SLAC for its support and the kind hospitality extended to them. 
This work is supported by the
US Department of Energy
and National Science Foundation, the
Natural Sciences and Engineering Research Council (Canada),
Institute of High Energy Physics (China), the
Commissariat \`a l'Energie Atomique and
Institut National de Physique Nucl\'eaire et de Physique des Particules
(France), the
Bundesministerium f\"ur Bildung und Forschung
(Germany), the
Istituto Nazionale di Fisica Nucleare (Italy),
the Research Council of Norway, the
Ministry of Science and Technology of the Russian Federation, and the
Particle Physics and Astronomy Research Council (United Kingdom). 
Individuals have received support from the Swiss 
National Science Foundation, the A. P. Sloan Foundation, 
the Research Corporation,
and the Alexander von Humboldt Foundation.


\begin{thebibliography}{99}
\bibitem{conjugate} 
The symbol \Bz\ refers 
to the $B_d$ meson; charge conjugate
modes are implied throughout this paper.
\end{thebibliography}
\end{document}